\documentclass[aps,prb,superscriptaddress,reprint]{revtex4-1}
\usepackage{amsmath}
\usepackage{amsfonts}
\usepackage{amssymb}
\usepackage{graphicx}
\usepackage{array}

\begin{document}

\title{Finite Temperature Behavior in the  Second Landau Level \\ of the Two-dimensional Electron Gas}

\author{V. Shingla}
\affiliation{Department of Physics and Astronomy, Purdue University, West Lafayette, IN 47907}
\author{E. Kleinbaum}\email[]{Present address: Department of Electrical Engineering, Princeton University,  
Princeton, NJ 08544}
\affiliation{Department of Physics and Astronomy, Purdue University, West Lafayette, IN 47907}
\author{A. Kumar}
\affiliation{Department of Physics, Monmouth College, Monmouth, IL 61462}
\author{L.N. Pfeiffer}
\affiliation{Department of Electrical Engineering, Princeton University, Princeton, NJ 08544}
\author{K.W. West}
\affiliation{Department of Electrical Engineering, Princeton University, Princeton, NJ 08544}
\author{G.A. Cs\'athy}
\affiliation{Department of Physics and Astronomy, Purdue University, West Lafayette, IN 47907}
\affiliation{Birck Nanotechnology Center, Purdue University, West Lafayette, IN 47907}

\date{\today}

\begin{abstract}

Reports of weak local minima in the magnetoresistance 
at $\nu=2+3/5$, $2+3/7$, $2+4/9$, $2+5/9$, $2+5/7$, and $2+5/8$ in second Landau level of the electron gas in GaAs/AlGaAs
left open the possibility of fractional quantum Hall states at these filling factors.
In a high quality sample we found that the magnetoresistance exhibits peculiar features 
near these filling factors of interest.
These features, however, cannot be associated with fractional quantum Hall states;
instead they originate from magnetoresistive fingerprints of the electronic bubble phases. 
We found only two exceptions: at $\nu=2+2/7$ and $2+5/7$ there is evidence for incipient
fractional quantum Hall states at intermediate temperatures. As the temperature is lowered,
these fractional quantum Hall states collapse due to a phase competition with bubble phases.

\end{abstract}
\maketitle

The two-dimensional electron gas subjected to a perpendicular magnetic field is a model system that supports
a large variety of electronic phases \cite{tsui,jainBook,willett,fogler,lilly99,du99,pan99,eisen02,xia04,kumar10,ethan15}. 
Many new phases were discovered
in high quality GaAs/AlGaAs heterostructures and this system continues to play
an important role in the study of these phases. Improvements in the material quality
of bilayer graphene \cite{gr1,gr2} and ZnO \cite{zno} offer a chance to study different realizations of these phases
in alternative hosts.

The most fascinating region of current interest of the electron gas in GaAs/AlGaAs
is the second orbital Landau level. Here we find numerous fractional quantum Hall states (FQHSs)
\cite{willett,pan99,eisen02,xia04,kumar10,ethan15}.
Several of these FQHSs are thought to have topological order and exotic quasiparticle excitations
which cannot be realized in the lowest Landau level \cite{tsui}.
The most well-known of these is the $\nu=5/2=2+1/2$
FQHS  \cite{willett,pan99}, which is believed to belong to the Pfaffian universality class and to host
Majorana-like excitations \cite{mr,zalat}. The $\nu=2+2/5$ is another FQHS of interest \cite{xia04} as
it is a candidate hosting Fibonacci anyons \cite{rr1,rr2}. In addition to FQHSs,
the second Landau level also supports a set of traditional
Landau phases with charge order. Examples are
the electronic bubble phases \cite{fogler,eisen02,deng12}, but under special circumstances the quantum Hall nematic 
may also develop \cite{nodar16,kate17}.
The region of the second Landau level, therefore, stands out among other 
Landau levels in a prominent display of phase competition between two classes of different phases:
FQHSs and charge ordered phases \cite{xia04}. 

In most experiments, data in the second Landau level of the highest quality samples
exhibit a consistent set of ground states: one typically observes fully developed
FQHSs at $\nu=2+1/2$, $2+1/3$, $2+2/3$, $2+1/5$, $2+4/5$, and  
%\cite{eisen02,xia04,kumar10,ethan15,engel05,marcus07,dean08,pan08,radu08,heiblum08,heiblum09,nuebler10,willett10,jing10,shay10,shay11,dolev11,jing11,nodar11,pan11,setup,
%deng12,deng12b,muraki12,willett13,chick13,muraki13,pinczuk13,ensslin13,dengAlloy,folk14,ensslin15,watson15,gervais15,folk16,lin16,smet17,qi17,gervais17,choi08,pan12,ensslin14},
$2+2/5$ and up to four bubble phases. %\cite{xia04,pan08,dean08,kumar10,nodar11,deng12b,watson15,qi17,choi08,pan12}.
In addition, in setups reaching the lowest temperatures, several developing FQHSs are observed at 
$\nu=2+3/8$ \cite{xia04,kumar10,ethan15,pan12,qi17,choi08,deng12b,pan08,nodar11,watson15},
$2+6/13$ \cite{kumar10,ethan15,pan12,qi17,deng12b,watson15}, 
$2+2/9$ \cite{deng12b}, 
$2+7/9$ \cite{ethan15,deng12b,pan08} 
and $2+2/7$ \cite{xia04,dean08,choi08,pan12}.
However, a careful inspection of the literature reveals that 
there are several additional magnetoresistance minima, such as the ones at
$\nu=2+5/8$, $2+5/7$ in Ref.\cite{pan12},
$\nu=2+3/5$, $2+3/7$, $2+5/7$, $2+4/9$, $2+5/9$, $2+5/8$ in Ref.\cite{choi08},
and $\nu=2+4/9$, $2+5/9$, $2+5/7$ in Ref.\cite{ensslin14}.
Even though these minima develop at filling factors compatible with FQHSs,
they could not be associated with FQHSs either because of lack of Hall data \cite{choi08,pan12}
or because the quantization of the Hall resistance was not consistent with that of a FQHS \cite{ensslin14}.
Furthermore, with the exception of $\nu=2+5/8$, at the filling factors of these additional minima other experiments report
bubble phases at either lower electron temperatures and/or in higher quality samples 
\cite{xia04,kumar10,ethan15,qi17,deng12b,pan08,nodar11}.
%Such a 

There may be several reasons for the development of these additional local minima in $R_{xx}$
in certain experiments but of bubble phases in others.
First, samples with different growth parameters have
different electron-electron interaction that may result in a drastically different set of ground states.
It is thus possible that, with improvement of sample quality,
the signatures seen in Refs\cite{pan12,choi08,ensslin14} develop into quantized FQHSs. 
Second, the available data may indicate a temperature-driven phase competition of FQHSs and bubble phases.
Indeed, there are well-known FQHSs present at
intermediate temperatures, which give way to a charge-ordered phase at the lowest accessible temperatures.
Examples of such FQHSs are at $\nu=1/7$ \cite{hiB1,hiB2} and $2/11$ FQHSs \cite{hiB2} 
in the lowest Landau level, $\nu=4+1/5$ and $4+4/5$ in the third Landau level \cite{gervaisB},
and $\nu=2+2/7$ in the second Landau level \cite{xia04}. Of these, the FQHSs observed at intermediate temperatures
in the second and third Landau levels turn into bubble phases as the temperature is lowered.

Here we examine whether the earlier seen minima in $R_{xx}$ that could not be associated with a FQHS
also develop in the second Landau level of a high quality GaAs/AlGaAs sample. 
We are interested in examining previously unavailable detailed temperature dependence
to observe phases at intermediate temperatures.
While in our sample we find peculiar features in the magnetoresistance in the vicinity of the filling factors
of interest $\nu=2+3/5, 2+3/7, 2+4/9, 2+5/9$, and $2+5/8$, we cannot 
associate FQHSs with these filling factors neither at the lowest nor at any finite temperatures.
We show that these features arise from the development of the magnetoresistive fingerprints of the bubble phases.
In contrast, at $\nu= 2+2/7$ and $2+5/7$ we observe incipient FQHSs at intermediate temperatures, which
yield to a bubble phase as the temperature is lowered further.
Such a study is timely, because of the conflicting results reported
in the second Landau level of the GaAs/AlGaAs system. Furthermore, our work
is expected to be relevant for studies of bilayer graphene, in which an increasing number of
FQHSs \cite{gr1,gr2} as well as of bubble phases have been recently reported \cite{cory}.
 
Our sample is a symmetrically doped
30~nm quantum well sample with electron density $n=3.0 \times 10^{11}$/cm$^2$ 
and mobility $\mu=32 \times 10^6$cm$^2$/Vs. 
Following the procedure described in the Supplement of Ref.\cite{dengAlloy},
the sample state was prepared by a low temperature illumination with a red light emitting diode.
Our sample is the same as the one used in Ref.\cite{kumar10};
data presented in Figs.1-4 are, however, from a different sample state preparation than those from Ref.\cite{kumar10}. 
The sample is mounted in a He$^3$ immersion cell which assures
electron thermalization to the base temperature of our dilution refrigerator and enables a 
convenient temperature measurement through quartz tuning fork viscometry \cite{setup}.
%The excitation current used for transport is 2~nA.

\begin{figure}[t]
 \includegraphics[width=.9\columnwidth]{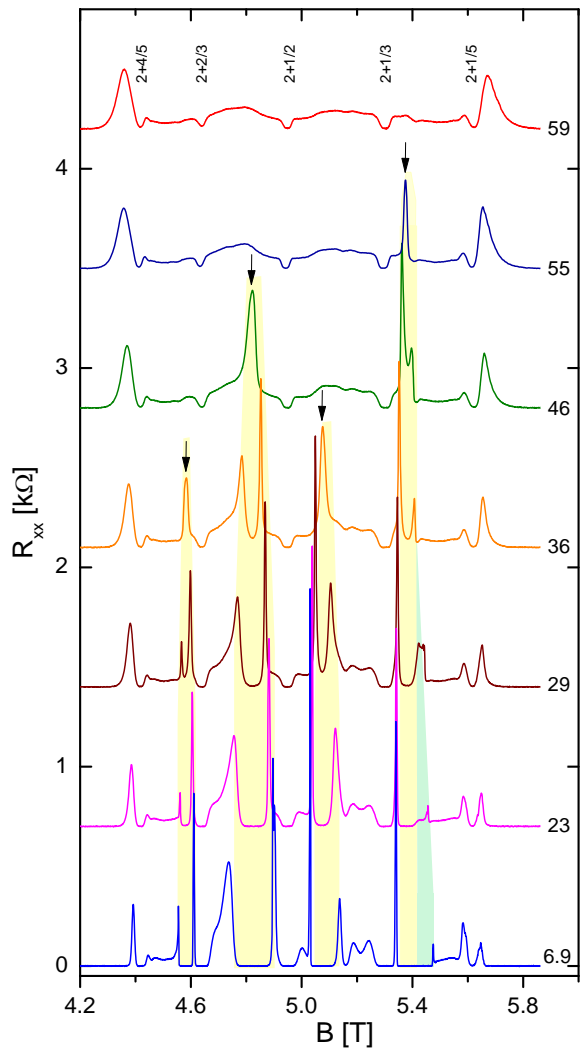}
 \caption{ Waterfall plot of the magnetoresistance in lower spin branch of the second Landau
 level ($2<\nu<3$). Filling factors of the five most prominent FQHSs are shown.
 The shaded areas mark the bubble phases present. Near $5.4$~T there are two different bubble phases present.
 Arrows indicate precursors of the bubble phases, i.e.
 transport features at the highest temperature that can still be associated with the bubbles.
 Numbers on the side show the measured temperatures in mK.}
 \end{figure} 

 \begin{figure*}
 \includegraphics[width=2\columnwidth]{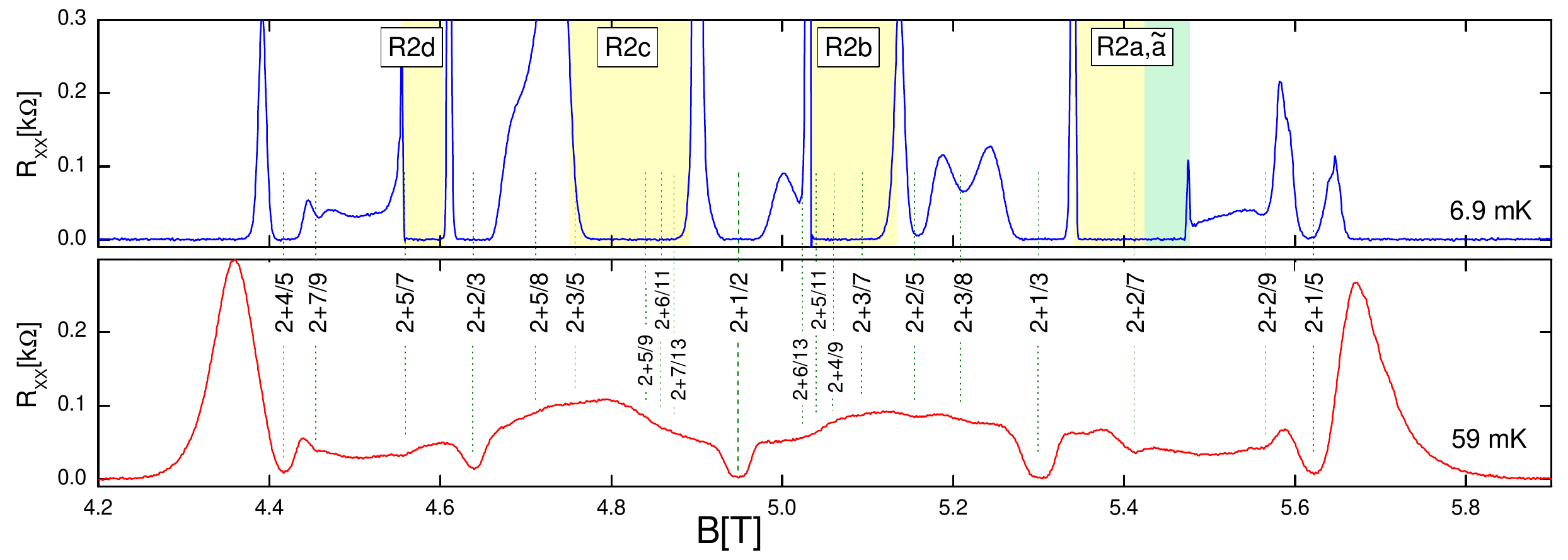}
 \caption{ A magnified view of the magnetoresistance at $2<\nu<3$ as
 measured at $T=59$~mK and $6.9$~mK. The various filling factors of interest
 are marked by vertical lines. The shaded areas are bubble phases.
 \label{Fig}}
 \end{figure*} 
  
Figure 1 captures the temperature evolution of magnetoresistance traces
in the second Landau level between $T=59$ and $6.9$~mK. 
We observe several FQHSs; the most prominent of these are
the ones at $\nu=2+1/2, 2+1/3, 2+2/3, 2+1/5$, and $2+4/5$. 
Traces of Fig.1 appear very different from
those measured in the lowest Landau level \cite{tsui} because of the presence of the reentrant
integer quantum Hall states  \cite{eisen02,deng12}. These reentrant states
are believed to be exotic electronic solids called the bubble phases \cite{fogler,lilly99,du99}.
The bubble phases we observe are marked by shading in Fig.1. At the lowest temperatures,
the bubble phases are signaled by a vanishing $R_{xx}$ and a Hall resistance quantized 
to either $h/2e^2$ or $h/3e^2$ (not shown) \cite{eisen02}.
Furthermore, the bubble phases are delimited by two distinct peaks in $R_{xx}$,
which can be seen near the edges of the shaded areas \cite{deng12}. The size of such
peaks may exceed $1.8$~k$\Omega$, hence they dominate the magnetoresistive landscape.
It was found that as the temperature is raised, the
two peaks delimiting a bubble phase first merge into a single peak, this single peak then
disappears as the temperature is increased further. 
Since these single peaks are the highest temperature signatures of the bubble phases,
they can be thought of as the precursors of the bubble phases.
In Fig.1 there are several examples marked by vertical arrows, such as the precursor peak at $B \simeq 5.37$~T in the $T=55$~mK trace.
At lower temperatures, near $B=5.4$~T there are two distinct
bubble phases, which will be discussed later.

A magnified view of the $T=6.9$ and $59$~mK traces is seen in Fig.2.  We singled out the $T=59$~mK
trace since this is the lowest temperature at which there are no discernible features of the bubble phases.
On this trace we marked several filling factors of interest:
the prominent FQHSs at $\nu=2+1/2, 2+1/3, 2+2/3, 2+1/5$, and $2+4/5$. 
Additional features are seen at several other filling factors. 
Some are relatively narrow depressions in $R_{xx}$, such as the 
ones at $\nu=2+2/5$, $2+2/7$, $2+2/9$, $2+7/9$, $2+5/7$ and $2+3/8$.
Other features are broader, such as the ones near $\nu=2+3/5$ and also in the vicinity of
$\nu=2+1/2$, on each side.
Of these features not all develop into a FQHS at $T=6.9$~mK. Indeed,
in the $T=6.9$~mK trace we identify fully developed FQHSs at
$\nu=2+1/2$, $2+1/3$, $2+2/3$, $2+2/5$ and less developed
FQHSs at $\nu=2+6/13$, $2+2/9$, $2+7/9$, and $2+3/8$. 
In the following we will examine the temperature dependence of the additional features of $R_{xx}$
shown in the $T=59$~mK trace of Fig.2. 
We will search, in particular, for signs of developing FQHSs which may be present at
intermediate temperatures, but which may not survive to the lowest accessible temperatures.

We first focus at filling factors related by particle-hole conjugation $\nu=2+3/5$ and $2+2/5$.
Interest in these quantum numbers stems from proposals and numerical evidence
that FQHSs here have a very special topological order
supporting non-Abelian anyons of the Fibonacci type \cite{rr1,rr2}.
Features in magnetotransport at these two filling factors were first found
and tentatively associated with FQHSs in Ref.\cite{pan99}. However, quantized Hall resistance was not observed;
the Hall resistance instead had features which were later attributed to the bubble phases.
A fully developed $\nu=2+2/5$ FQHS was observed in Ref.\cite{xia04} and it is now routinely
measured  \cite{xia04,kumar10,ethan15,pan12,qi17,dean08,choi08,deng12b,pan08,nodar11,watson15}.
In contrast to observations at $\nu=2+2/5$, 
at $\nu=2+3/5$ a FQHS was not detected in most experiments
\cite{xia04,kumar10,ethan15,pan12,qi17,dean08,deng12b,pan08,nodar11,watson15}. The filling factor
$\nu=2+3/5$ often falls very close to the bubble phase $R2c$ instead.
We are aware of only one work, in which a concave feature in $R_{xx}$ was seen at
$\nu=2+3/5$ \cite{choi08} at a temperature $T=36$~mK. We note that results in
wide quantum wells are qualitatively different; we defer discussing these results to a later paragraph.

\begin{figure}[b]
 \includegraphics[width=1\columnwidth]{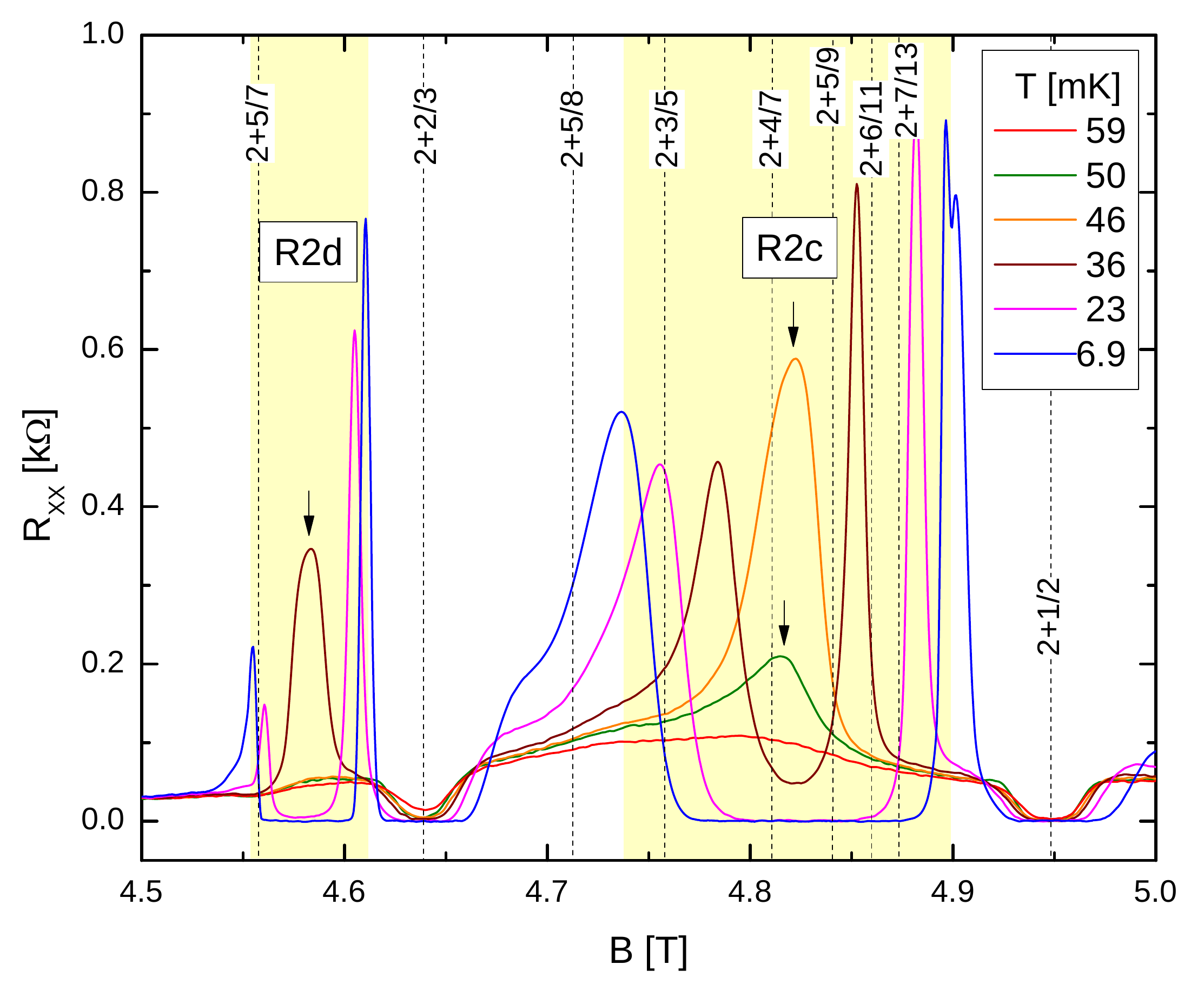}
 \caption{Details of the $T$-dependence of the magnetoresistance at filling factors
 less than $2+1/2$. Vertical arrows mark the precursors of the bubble phases $R2c$ and $R2d$.
  }
 \end{figure} 
 
As seen in Fig.2, our $T=6.9$~mK trace in the vicinity of $\nu=2+3/5$
is similar to that seen in 
Refs.\cite{xia04,kumar10,ethan15,qi17,deng12b,pan08,nodar11,watson15}, as
$\nu=2+3/5$ falls near a bubble phase. The $T=59$~mK trace of Fig.2,
however, is not perfectly smooth and it has a slight curvature at $\nu=2+3/5$. 
In order to establish whether this feature develops into
a FQHS at intermediate temperatures, in Fig.3 we examine data at intermediate temperatures.
We notice, that at $T=50$~mK a resistance peak appears near $B=4.82$~T.
This peak was associated with the bubble phase and can be thought of as the precursor of the bubble
phase labeled $R2c$. As the temperature is lowered to $T=46$~mK this peak grows, then at $T=36$~mK
it splits into two peaks, giving way to a pronounced resistance minimum between them.
Inspecting the data shown in Fig.3 we see that the precursor peaks of the bubble phase at $T=46$ and $50$~mK 
have a concave curvature on both sides, including one near $\nu=2+3/5$. These concave
features, however, cannot be associated with a developing FQHS. 
We thus conclude that, in spite of a fully developed FQHS at $\nu=2+2/5$, 
in our sample we do not observe signs of fractional correlations at $\nu=2+3/5$.
Recent theory work has considerably strengthened the case
for a Read-Rezayi state at $\nu=2+2/5$  \cite{papic,sheng,peterson} and
has addressed the experimentally observed asymmetry between $\nu=2+2/5$ and $2+3/5$. 
Two causes for the suppression of fractional correlations at $\nu=2+3/5$ were identified:
an enhanced Landau level mixing \cite{peterson} and an extremely close energetic competition
between the Read-Rezayi state and the bubble phase \cite{papic}. 
While in experiments both effects are likely to be present, results of Ref.\cite{papic} are particularly
relevant for our observations.

 \begin{figure}[t]
 \includegraphics[width=1\columnwidth]{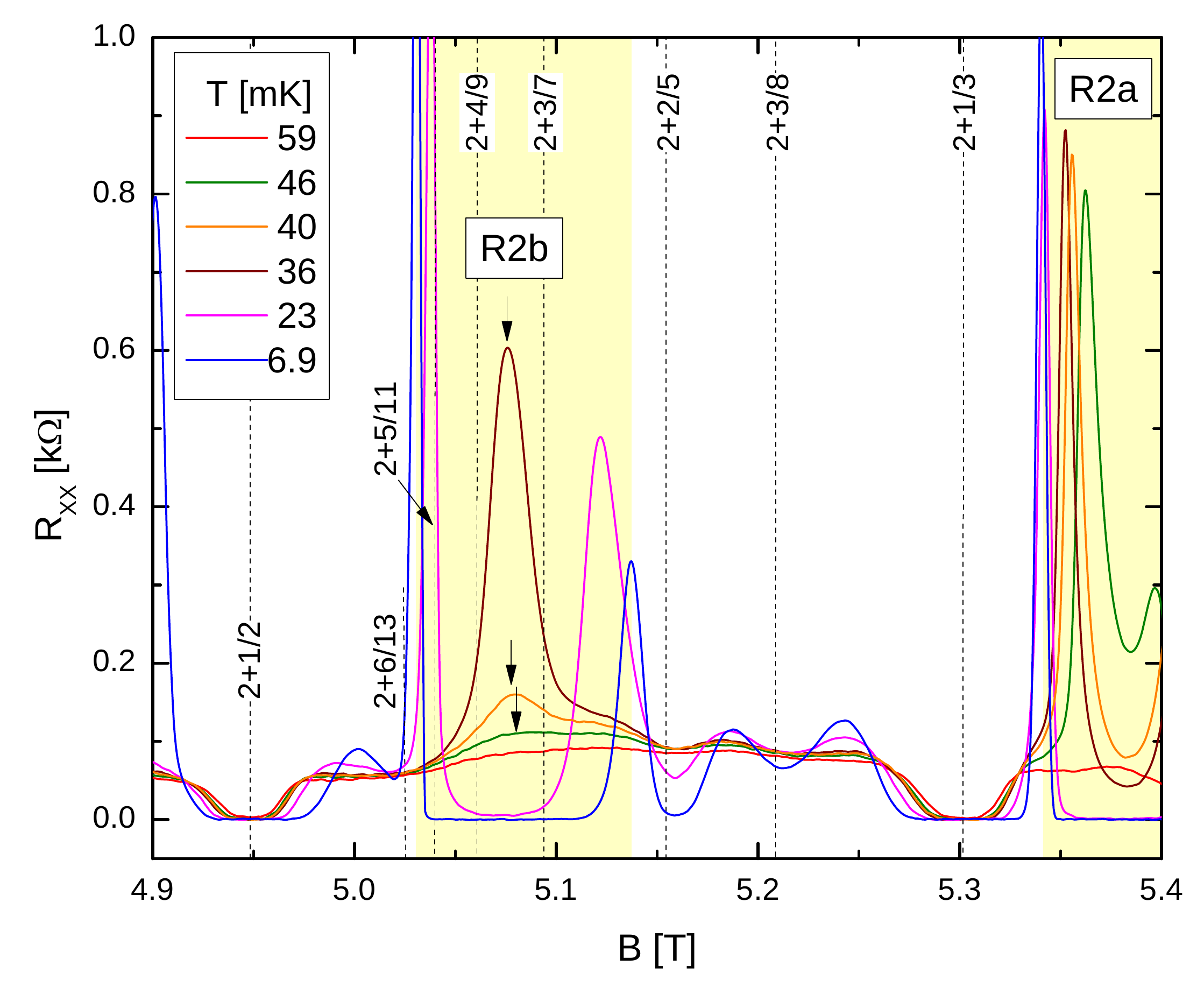}
 \caption{ Details of the $T$-dependence of the magnetoresistance at filling factors
 larger than $2+1/2$. Vertical arrows mark the precursors of the
 bubble phase $R2b$.
 }
 \end{figure} 
 
We note that different physics may be at play at $\nu=2+2/5$ and $2+3/5$
in GaAs/AlGaAs electron gases in which two electric subbands are occupied, such as electron gases confined to
wide quantum wells. It was shown that, in contrast to samples with a single
subband populated, in these systems the $\nu=2+2/5$ and $2+3/5$ filling factors
can be reached while the chemical potential is in the lowest Landau level \cite{shay10,shay11}. 
Under such circumstances, FQHSs have been observed both at $\nu=2+2/5$ and $2+3/5$.
These FQHSs, however, inherit the Laughlin-Jain correlations of the $\nu=2/5$ and $3/5$
FQHSs commonly observed in the lowest Landau level. Furthermore, under such circumstances
no bubble phases were observed, therefore a competition between FQHSs and bubble phases
does not occur \cite{shay10,shay11}.

%Sheng - Coulomb ground state, limit of exact particle-hole symmetry.
% at 13/5 density-matrix renormalization group calculations, on sphere N_e=36 and infinite cylinder
%establish incompressibility, and that the entanglement spectrum fits the conformal field theory of the
%parafermionic Read-Rezayi edge.  
%Finite-size scaling at both 12/5 and 13/5 confirms that the ground state with topological shift of the Read-Rezayi 
%parafermion state and its perticle-hole conjugate are energetically favored in the thermodynamic limit.
%Results at both 12/5  and 13/5 for the entanglement spectrum and the value of the 
% topological entanglement entropy show that the edge structure 
%and bulk quasiparticle statistics are consistent with the prediction based on the RR3 state.

%Peterson- exact diagonalization including
%finite well width and Landau-level mixing. We find that Landau-level mixing suppresses the ν = 13/5 fractional
%quantum Hall effect relative to ν = 12/5. Our results provide a possible explanation for the experimental absence of the 13/5
%fractional quantum Hall state as caused by Landau-level mixing effects.

%Papic- infinite density-matrix renormalization group calculations point out extremely close energetic competition
%between the Read-Rezayi phase and a reentrant integer quantum Hall phase. This competition suggests that
%even small particle-hole symmetry breaking perturbations can explain the experimentally observed asymmetry
%between ν = 12/5 and 13/5. We find that at ν = 12/5 Landau level mixing favors the Read-Rezayi phase over the reentrant phase.

  \begin{figure}[t]
 \includegraphics[width=1\columnwidth]{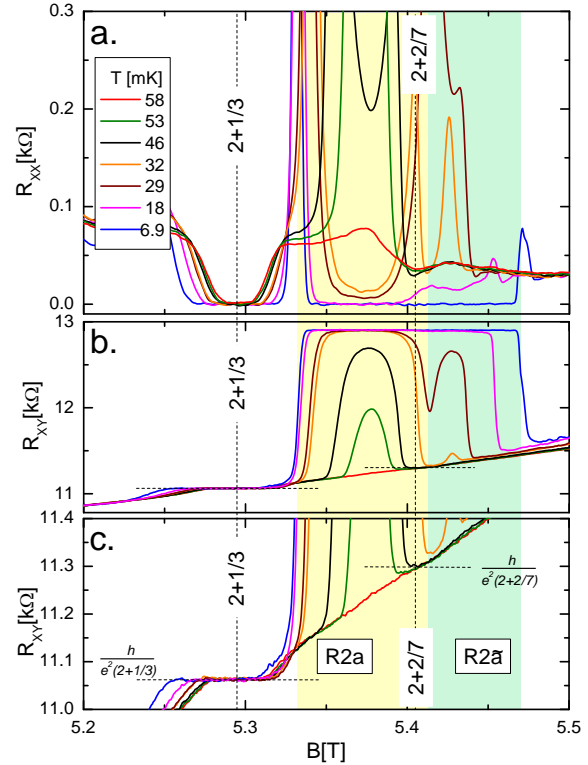}
 \caption{Temperature dependence of the magnetoresistance in the vicinity of $\nu=2+2/7$ (panel a).
 The Hall resistance for the same range of filling factors (panel b) and a magnified view of the
 Hall resistance (panel c).  Shading marks 
 the bubble phases $R2a$ and $R2\tilde{a}$ at $6.9$~mK. These bubble phases are
 separated by a deep minimum in $R_{xy}$ seen at $T=29$ and 32~mK, as shown in panel b.
 Quantization of the Hall resistance
  at $\nu=2+1/3$ and at $2+2/7$ is marked by horizontal dotted lines.
  }
 \end{figure} 
 
We now examine the range of filling factors from $\nu=2+1/2$ to $2+2/5$.
There are several references that report either a bubble phase \cite{eisen02,xia04,pan08,nodar11,deng12,deng12b,kumar10,chick13} 
or a precursor to the bubble phase in this region \cite{dean08,nuebler10}. The bubble phase in this range of fillings
is labeled $R2b$ in Fig.4. Signatures of fractional correlations in this region
were reported only in a handful of experiments.
A FQHS was reported at $\nu=2+6/13$ in Ref.\cite{kumar10};
this state has since been seen in other high mobility samples \cite{deng12b,pan12,qi17,watson15}. 
In these experiments no other FQHSs were observed in the $2+2/5 < \nu < 2+1/2$
region \cite{kumar10,deng12b,pan12,qi17}.
In contrast, local minima were reported at $\nu=2+3/7$ and
$2+4/9$, but the bubble phase $R2b$ was not observed in Ref.\cite{choi08}.
In addition, in Ref.\cite{ensslin14}, a local minimum in $R_{xx}$ was also observed at
$\nu=2+4/9$, although the Hall resistance at this filling factor was not quantized.
In our sample we observe a developing FQHS at $\nu=2+6/13$.
Furthermore, at $T=36, 40, 46$~mK in our data we observe precursor peaks associated 
with the bubble phase $R2b$. These precursor peaks exhibit a concave curvature
on both of their sides, near $\nu=2+4/9$ and $2+3/7$. However, the concave features in our sample
in the vicinity of these two filling factors cannot be associated with a developing FQHS.
We thus conclude, that in our sample there is no evidence of FQHSs
at $\nu=2+3/7, 2+4/9, 2+5/11$ at any of the temperatures examined.
Fig.3 shows that a similar conclusion can be reached at filling factors
$\nu=2+7/13$, $2+6/11$, $2+5/9$, $2+4/7$, $2+3/5$, and $2+5/8$;
of these filling factors a local minimum in $R_{xx}$ was seen at $\nu=2+5/9$ in Refs.\cite{choi08,ensslin14}
and at $\nu=2+5/8$ in Ref.\cite{pan12}. 
%The feature in the magnetoresistance at $\nu=2+5/8$ in our sample is not associated with the precursor of
%$R2c$; it is instead a curvature developing on one of the peaks delimiting the $R2c$ bubble.
We thus found that curvatures in the magnetoresistance of our sample at the filling factors enumerated above
cannot be associated with incipient fractional quantum Hall states;
instead they originate from magnetoresistive fingerprints of the electronic bubble phases.

In contrast to the behavior of the magnetoresistance at the filling factors
discussed above, that at $\nu=2+2/7$ is quite different. As discussed in Ref.\cite{xia04},
with the lowering of the temperature, $R_{xx}$ at this filling factor drops and
$R_{xy}$ approaches full quantization. Our sample shows a similar behavior.
Fig.5a shows that at $T=46$~mK and $\nu=2+2/7$, $R_{xx}$  reaches its lowest value.
As shown in Fig.5b and Fig.5c, at this temperature and filling $R_{xy}$ becomes equal to $h/(2+2/7)e^2$ within our measurement error.
In contrast to Refs.\cite{xia04,dean08,choi08,pan12}, transport at the lowest temperature in our sample at $\nu=2+2/7$
exhibits a fully developed reentrant insulator, i.e. $R_{xx}=0$ and $R_{xy}=h/2e^2$.
As already reported, near $\nu=2+2/7$ there are two distinct bubble phases \cite{xia04,deng12}, 
labeled $R2a$ and $R2\tilde{a}$ in Fig.5. Shading in this figure denotes the stability range
of these bubble phases at $6.9$~mK; the two different bubbles are delimited by the deep minimum in $R_{xy}$
shown in Fig.5b.
It is interesting to note that this deep minimum in $R_{xy}$ is close to, but not
at $\nu=2+2/7$. We find a similar behavior at the related filling factor $\nu=2+5/7$.
Indeed, in Fig.2 we observe a conspicuous minimum in $R_{xx}$ at $T=59$~mK at this filling factor.
Such a local minimum was also observed in Ref.\cite{ensslin14} and it
may indicate developing fractional correlations. 
However, as shown in Fig.3, this minimum at $\nu=2+5/7$ disappears with the lowering of the temperature and 
the $R2d$ bubble phase prevails.

Our observations are expected to be relevant for the two-dimensional electron gas confined to bilayer graphene.
Improvements in the quality of this system revealed an increasing number of FQHSs, including
even denominator FQHSs \cite{gr1,gr2}. Details, such as the
nature of the wavefunction in the $N=1$ Landau level and the presence of the valley degree of freedom in bilayer
graphene, result in differences in the physics, when compared to that in the GaAs/AlGaAs system \cite{gr1,gr2}. 
Nonetheless, in addition to FQHSs, the most recent measurements in bilayer graphene also reveal
reentrant integer quantum Hall effect commonly associated with bubble phases \cite{cory}. 
Bilayer graphene is thus expected to display phase competition between FQHSs
and bubble phases similar to that seen in the GaAs/AlGaAs system.

To conclude, the development of precursors of the electronic bubble phases in the second Landau level of 
two-dimensional electron gases confers strong concave features to the magnetoresistance.
In the high quality sample we studied, these concave features 
cannot be associated with any developing FQHSs. 
In contrast, the local minima present in the magnetoresistance
at intermediate temperatures developing at $\nu=2+2/7$ and $2+5/7$ 
are interpreted as being due to incipient fractional quantum Hall states. 
However, as the temperature is lowered, these 
incipient FQHSs collapse due to a phase competition with an electronic bubble phase.

Measurements at Purdue were supported by the NSF Grant No. DMR 1505866. Sample growth effort
of L.N.P. and K.W.W. of Princeton University was supported by the Gordon and Betty Moore Foundation
Grant No. GBMF 4420, and the National Science Foundation MRSEC Grant No. DMR-1420541.

\end{document}